\documentclass[twocolumn,preprintnumbers]{revtex4}
\usepackage[dvipdfmx]{graphicx,color}
\usepackage{amsmath,amssymb}
\newcommand{\ket}[1]{| #1 \rangle}

\newcommand{\He}{^6{\rm He}}
\newcommand{\Be}{^{10}{\rm Be}}
\newcommand{\Li}{^{9}{\rm Li}}

\newcommand{\bvec}{\boldsymbol}

\begin{document}
\title{Monopole transitions to cluster states in $^{10}$Be and $^9$Li}
\author{Yoshiko Kanada-En'yo}
\affiliation{Department of Physics, Kyoto University, Kyoto 606-8502, Japan}

\begin{abstract}
Isoscalar monopole transitions from the ground states 
to cluster states in $\Be$ and $\Li$ 
are investigated with $\He+\alpha$ and $\He+t$ cluster models, respectively.
In $\Be$, significant monopole strengths to 
$\He+\alpha$ cluster resonances of $\Be(0^+_{3,4})$ 
above the $\alpha$-decay threshold 
are obtained, whereas 
those to $\He+t$ cluster resonances in $^9$Li are not enhanced because of 
the large fragmentation of the strengths in the corresponding energy region.
The monopole transition to $\Be(0^+_2)$ having the molecular orbital 
structure is relatively weak compared with those to $\He+\alpha$ 
cluster resonances.
 Monopole strength distributions do not directly 
correspond to distributions of 
$\He(0^+)+\alpha$ and $\He(0+)+t$ components but 
they reflect component of the deformed $\He$
cluster with a specific orientation, 
which is originally embedded in the ground state.
\end{abstract}
\maketitle

\noindent

\section{Introduction} \label{ref:introduction}
In this decade, various exotic cluster states have
been discovered in neutron-rich nuclei.
Neutron-rich Be isotopes are typical examples, in which 
a variety of cluster structures appear in the ground and excited states (for example, Refs.~\cite{Oertzen-rev,KanadaEn'yo:2001qw,KanadaEn'yo:2012bj,Ito2014-rev}
and references therein). 
In Be isotopes, low-lying states are understood by a molecular orbital picture where 
valence neutrons in molecular orbitals around a $2\alpha$ core are considered \cite{Oertzen-rev,KanadaEn'yo:2012bj,Ito2014-rev,OKABE,SEYA,OERTZENa,OERTZENb,Arai:1996dq,Dote:1997zz,Ogawa:1998et,KanadaEn'yo:1999ub,
Itagaki:1999vm,Itagaki:2000nn,Itagaki:2001az,
Descouvemont01,Descouvemont02,KanadaEn'yo:2002rh,KanadaEn'yo:2003ue,Ito:2003px,oertzen03-rev,Arai:2004yf,
Ito:2005yy,Ito:2008zza,Dufour10,Ito:2011zza,Ito:2012zza,KanadaEn'yo:2012rm}.
In highly excited states 
above the He+He threshold energy, dinuclear-type 
He+He resonances (cluster resonances)
are expected to appear as suggested in $^{10}$Be and $^{12}$Be
\cite{,Oertzen-rev,KanadaEn'yo:2012bj,Ito2014-rev,Descouvemont02,
KanadaEn'yo:2003ue,Ito:2003px,oertzen03-rev,Arai:2004yf,Ito:2005yy,Ito:2008zza,Dufour10,Ito:2011zza,Ito:2012zza,KanadaEn'yo:2012rm,Fujimura:1999zz,KanadaEn'yo:2002ay,
Hamada:1994zz,Soic:1995av,Curtis:2001sd,Liendo:2002gx,Fletcher:2003wk,Curtis:2004wr,
Ahmed:2004th,Millin05,Freer:2006zz,
Bohlen:2007qx,Curtis:2009zz,Suzuki:2013mga,bohlen98,bohlen02,korsheninnikov95,
Freer:1999zz,Freer:2001ef,SAITO04,Yang:2014kxa}.

The coexistence of molecular orbital structures and cluster resonances 
in $^{12}$Be has been intensively studied by experimental and theoretical works.
The ground state of $^{12}$Be is a largely deformed intruder state, in which  
two neutrons occupy a longitudinal molecular orbital so-called the $\sigma$ orbital 
around the $2\alpha$ core. Highly excited states observed by 
$^6{\rm He}+^6{\rm He}$ and $^8{\rm He}+^4{\rm He}$ decays above the
threshold energies are regarded as cluster resonances \cite{Freer:1999zz,Freer:2001ef,SAITO04,Yang:2014kxa}. 
In theoretical studies with the generalized two-center cluster model (GTCM),  
Ito {\it et al.} predicted that $^8{\rm He}+^4{\rm He}$, $^6{\rm He}+^6{\rm He}$,
$^7{\rm He}+^5{\rm He}$ cluster resonances appear  in the energy  region a few MeV above 
the threshold energies \cite{Ito2014-rev,Ito:2008zza,Ito:2011zza,Ito:2012zza}. They discussed monopole transitions 
from the ground state to the excited states and showed that the monopole strength 
to the $^8{\rm He}+^4{\rm He}$ cluster resonance is strongly enhanced.
It means that monopole excitations can be a good probe to experimentally 
observe cluster resonances.  

The coexistence of molecular orbital structures and cluster resonances 
has been also investigated for $\Be$. Theoretical works predicted 
the $K^\pi=0^+_2$ band constructed by a largely deformed state with
a molecular orbital structure having two $\sigma$-orbital neutrons around
the developed $2\alpha$ core.
In the experimental energy levels, 
the $0^+_2$ state at 6.18 MeV, the  
$2^+$ state at 7.54 MeV, and the $4^+$ state at 10.2 MeV 
are assigned to the $K^\pi=0^+_2$ band \cite{Soic:1995av,Liendo:2002gx,Millin05,Freer:2006zz},
though the spin and parity of the 10.2 MeV state have not been established yet \cite{Curtis:2004wr}. 
Above the $K^\pi=0^+_2$ band, $\He+\alpha$ cluster resonances have been theoretically predicted
\cite{Kobayashi:2012di,Ito2014-rev}, however,  
there is as yet no experimental evidence of cluster resonances above the $\He+\alpha$ decay threshold in $\Be$.
There is only an experimental report of a broad resonance in $^{10}{\rm B}$, which is regarded as the 
mirror state of a $\He+\alpha$ cluster resonance \cite{Kuchera:2011ax}. 

In analogy to $\Be$, cluster states in excited states of 
$\Li$ have been theoretically studied using a $\He+t$ cluster model 
by the author and her collaborators \cite{KanadaEn'yo:2011nc}. They predicted 
$\He(0^+)+t$ cluster resonances in highly excited states above the $t$-decay 
threshold, which can be analogous to $\He(0^+)+\alpha$ cluster resonances in $\Be$.
It was also shown that molecular orbital structures do not appear in $\Li$ 
because molecular orbitals are unfavored around the
asymmetric core of $\alpha+t$, differently from $\Be$ having the symmetric 
core of $2\alpha$. 

In the present paper, I investigate monopole excitations from the ground states
to excited $\Be(0^+)$ and $\Li(3/2^-)$ states. Attention is focused on monopole strengths
to cluster resonances to answer a
question whether monopole strengths can be probes to observe cluster 
resonances.
For this aim, I adopt the generator coordinate method (GCM) \cite{GCM} 
of the $\He+\alpha$ and $\He+t$ cluster models 
performed to investigate cluster states in $\Li$ and $\Be$ in the previous work \cite{KanadaEn'yo:2011nc}. 
I reanalyze $\Li(3/2^-)$ and $\Be(0^+)$ states 
while focusing on monopole excitations.
The method has been proved to describe well experimental properties of the ground 
and the second $0^+$ bands in $\Be$. 
In the calculation, resonance states are obtained in a bound state approximation. 
I estimate cluster decay widths of the cluster resonances from approximated 
reduced width amplitudes at channel radii, and also evaluate them   
by changing the size of the box boundary. The monopole strengths for transitions from 
the ground state to excited states are investigated. 
Cluster components in the ground and excited states are calculated, and their relation to 
monopole excitations is discussed.

This paper is organized as follows.
In section \ref{sec:formulation}, I explain the formulation of 
the present calculation. I show the calculated results in section \ref{sec:results}
and give discussions of cluster structures and monopole excitations in section \ref{sec:discussions}. 
Finally a summary and an outlook are given in section \ref{sec:summary}.

\section{Formulation}\label{sec:formulation}

\subsection{$\He+\alpha(t)$ cluster wave functions}
The Bloch-Brink (BB) wave functions \cite{brink66} 
of $\He+\alpha$ and $\He+t$ cluster wave functions are used as done in the previous work
\cite{KanadaEn'yo:2011nc}.
The $\He$ and $\alpha$($t$) cluster wave functions are written 
by harmonic oscillator (ho) shell-model wave functions 
localized at $\bvec{S}_1=(0,0,-\frac{A_2}{A}D)$ and $\bvec{S}_2=(0,0,+\frac{A_1}{A}D)$, respectively. 
Here $A_1$ and $A_2$ are the mass numbers of two clusters and $A$ is the total
mass number $A=A_1+A_2$.
$D$ indicates the
distance parameter, which is treated as the generator coordinate in the superposition of basis wave functions. 
A common width parameter $\nu=1/2b^2=0.235$ fm$^{-2}$ is used 
for $\He$, $\alpha$, and $t$ clusters.

The $\alpha$ and $t$ clusters are expressed by $(0s)^{2}_\pi(0s)_\nu^2$ and $(0s)_\pi(0s)_\nu^2$ configurations,
respectively.
For the $\He$ cluster, $p$-shell configurations of two valence neutrons around 
an $\alpha$ cluster, $(0s)_\pi^2 (0s)_\nu^2 (0p)_\nu^2$, are used. 
To express $p$-shell configurations, I use the hybrid model space of basis wave functions
combining  $ls$ and $jj$ coupling schemes as done in the previous work.
For the configurations favored in the $ls$ coupling scheme, 
$\ket{p_z, n\uparrow}\ket{p_z, n\downarrow}$ and 
its rotated configurations are used to take into account $\He(0^+,2^+)$ states 
with the two-neutron intrinsic spin $S_{12}=0$. For those favored in the $jj$ coupling scheme, 
$\ket{p_{(+)}, n\uparrow}\ket{p_{(-)}, n\downarrow}$ and its rotated 
configurations are adopted to take into account 
$\He(0^+,2^+)$ states in the $p_{3/2}^2$ configurations. 
Here, $p_{(+)}, p_z, p_{(-)}$ stand for the ho $p$ orbits with 
$l_z=+1,0,-1$, respectively  ($l_z$ is the $z$-component 
of the orbital angular momentum $\bvec{l}$). 
Note that the hybrid model space of these $ls$ coupling and $jj$ coupling configurations 
for the  $\He$ cluster 
is equivalent to the full model space of $p$-shell configurations for $\He(0^+,2^+)$.

\begin{figure}[th]
\begin{center}
\includegraphics[width=7.5cm]{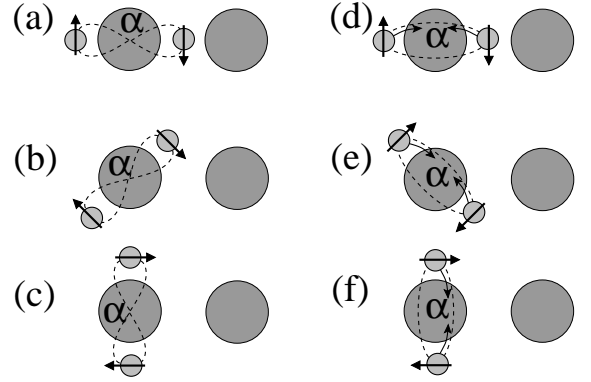} 	
\end{center}
\caption{Schematic figures for configurations $(a)$-$(f)$ of the $^6$He cluster in the
$^6$He+$t$ and $^6$He+$\alpha$ cluster models. Details are described in the text.
}
\label{fig:he6-t}
\end{figure}

The $\He+\alpha$ and $\He+t$ cluster wave functions projected onto parity and 
total-angular-momentum eigen states are written as,
\begin{eqnarray}
&&P^{J\pi}_{MK} \ket{\Phi_{\tau}(D)}=P^{J\pi}_{MK} {\cal A}\left\{ \ket{\psi_{1\tau}(\bvec{S}_1}\ket{\psi_{2\tau}(\bvec{S}_1)} \right. \nonumber\\
&&\times \ket{\phi(\bvec{S}_1)p\uparrow}
\ket{\phi(\bvec{S}_1) p\downarrow}
\ket{\phi(\bvec{S}_1) n\uparrow}
\ket{\phi(\bvec{S}_1) n\downarrow}  \nonumber\\ 
&& \times \left. \ket{\phi(\bvec{S}_2) p\uparrow}
 \ket{\phi(\bvec{S}_2)p\downarrow}
\ket{\phi(\bvec{S}_2)n\uparrow}
\ket{\phi(\bvec{S}_2)n\downarrow} \right\}
\end{eqnarray}
and
\begin{eqnarray}
&&P^{J\pi}_{MK} \ket{\Phi_{\tau}(D)}=P^{J\pi}_{MK} {\cal A}\left\{ \ket{\psi_{1\tau}(\bvec{S}_1}\ket{\psi_{2\tau}(\bvec{S}_1)} \right. \nonumber\\
&&\times \ket{\phi(\bvec{S}_1)p\uparrow}
\ket{\phi(\bvec{S}_1) p\downarrow}
\ket{\phi(\bvec{S}_1) n\uparrow}
\ket{\phi(\bvec{S}_1) n\downarrow}  \nonumber\\ 
&& \times \left. \ket{\phi(\bvec{S}_2) p\uparrow}
\ket{\phi(\bvec{S}_2)n\uparrow}
\ket{\phi(\bvec{S}_2)n\downarrow} \right\}.
\end{eqnarray}
Here $\phi(\bvec{S}_i)$ is the $0s$ wave function shifted to the position $\bvec{S}_i$.
$\ket{\psi_{1\tau}(\bvec{S}_1)}$ and $\ket{\psi_{2\tau}(\bvec{S}_1)}$
indicate $p$-shell orbits for neutron configurations 
labeled by $\tau=\{a,b,c,d,e,f \}$ of the $\He$ cluster
shifted to $\bvec{S}_1$.
Schematic figures for the configurations $\tau=\{a,b,c,d,e,f \}$
are illustrated in Fig.~\ref{fig:he6-t}.

The $ls$ coupling configurations are given by 
the configurations $\tau=a$, $b$, and $c$, 
in which two neutron orbits are written by rotated configurations 
of  $\ket{p_y, n\uparrow}_{\bvec{S}_1}\ket{p_y, n\downarrow}_{\bvec{S}_1}$ as 
\begin{eqnarray}
&\ket{\psi_{1\tau}(\bvec{S}_1)}=\hat R_{x,\bvec{S}_1}(\theta)\ket{p_y, n\uparrow}_{\bvec{S}_1},\\
&\ket{\psi_{2\tau}(\bvec{S}_1)}=\hat R_{x,\bvec{S}_1}(\theta)\ket{p_y, n\downarrow}_{\bvec{S}_1},
\end{eqnarray}
with the rotation angle $\theta=\pi/2$, $\theta=\pi/4$, and  $\theta=0$, respectively. 
Here $\hat R_{x,\bvec{S}_1}(\theta)$ is the rotation operator around the $x$-oriented 
axis passing through $\bvec{S}_1$.

The $p_{3/2}^2$ configurations in the $jj$ coupling scheme 
are described by the configurations $\tau=d$, $e$, and $f$, 
in which two neutron orbits are given by rotated configurations of 
$\ket{p_{(+)}, n\uparrow}_{\bvec{S}_1}\ket{p_{(-)}, n\downarrow}_{\bvec{S}_1}$
 as 
\begin{eqnarray}
&\ket{\psi_{1\tau}(\bvec{S}_1)}=\hat R_{x,\bvec{S}_1}(\theta)\ket{p_{(+)}, n\uparrow}_{\bvec{S}_1},\\
&\ket{\psi_{2\tau}(\bvec{S}_1)}=\hat R_{x,\bvec{S}_1}(\theta)\ket{p_{(-)}, n\downarrow}_{\bvec{S}_1},
\end{eqnarray}
where $\theta=\pi/2$, $\theta=\pi/4$, and  $\theta=0$ are chosen for 
$d$, $e$, and $f$, respectively.

In the GCM calculation, the $\He+\alpha(t)$ cluster wave functions are superposed as
\begin{equation} \label{eq:gcm}
\ket{\Psi^{J^\pi_k}_M}=\sum_D \sum_{\tau,K} c^{(J^\pi_k)}_{D,\tau, K} P^{J\pi}_{MK} \ket{\Phi_{\tau}(D)}, 
\end{equation}
where the coefficients $c^{(J^\pi_k)}_{D,\tau, K}$ are determined by diagonalizing norm and Hamiltonian matrices. 
I use the generator coordinate $D \le D_{\rm max}$ and get wave functions for resonances states as bound-state solutions.

The present model space given by the $K$-projected wave functions of six configurations 
fully covers all $p$-shell configurations of $0^+$ and $2^+$ states of the $\He$ cluster located at the inter-cluster 
distance $D$.
A single configuration has the deformed $\He$ cluster with a specific orientation and indicates  
a strong-coupling cluster structure, in which angular momenta of clusters and that of the inter-cluster motion are
strongly coupled. 
At a moderate distance $D$, the configuration $(a)$ having two neutrons in the longitudinal direction
approximately corresponds to the molecular orbital $\sigma^2$ structure
because of the antisymmetrization effect, which is the dominant component of the $\Be(0^+_2)$ state 
consistently with preceding works \cite{SEYA,OERTZENa,OERTZENb,Arai:1996dq,Dote:1997zz,Ogawa:1998et,
KanadaEn'yo:1999ub,Itagaki:1999vm,Itagaki:2000nn,Itagaki:2001az,Descouvemont02,Ito:2003px,Ito:2005yy,Ito2014-rev}. 
On the other hand, in the asymptotic region of a large distance $D$, the system 
goes to a weak-coupling cluster state, in which the $^6$He subsystem becomes its energy eigen state
with a certain spin $I$.
In the present framework, the transition between strong-coupling 
to weak-coupling cluster structures 
is taken into account by the 
linear combination of configurations $(a)$-$(f)$ projected to total angular momentum eigen states. 
The present expression is useful to analyze cluster structures of $\He+\alpha(t)$ in $\Be$($\Li$), in particular, 
in a strong-coupling regime.

Transition between strong-coupling and weak-coupling cluster structures of $^{10}$Be was nicely presented by Ito {\it et al.} \cite{Ito:2003px,Ito:2005yy} using the GTCM, in which 
molecular orbital configurations are fully taken into account 
by channel coupling (configuration mixing) of the $^5$He+$^5$He channel 
with $^6$He+$\alpha$ channel. 
In the present work, I omit the coupling with the $^5$He+$^5$He channel
to save the number of basis wave functions.
In spite of omitting the $^5$He+$^5$He configurations, the present method works 
well to describe experimental 
energy spectra of the $K=0^+_1$,
$K=2^+_1$, and $K=0^+_2$ bands of $^{10}$Be as already shown in the previous work.

In the practical calculation, I express a configuration 
of the $\He+\alpha(t)$ wave functions with 
a single AMD wave function which is given by a Slater determinant of single-particle Gaussian wave packets. General form of the 
AMD wave functions is described, for example, 
in Refs.~\cite{KanadaEn'yo:2001qw,KanadaEn'yo:2012bj,KanadaEnyo:1995tb}.

\subsection{Isoscalar monopole transitions}
The isoscalar monopole (ISM) operator ${\cal M}(IS0)$ 
is defined as
\begin{eqnarray}
{\cal M}(IS0)&=&\sum_i  (\bvec{r}_i-\bvec{R})^2,
\end{eqnarray}
where $\bvec{r}_i$ is the $i$th nucleon coordinate and 
$\bvec{R}$ is the center of mass coordinate $\bvec{R}\equiv \sum_i\bvec{r}_i/A$. 
The ISM strength from the ground state
to an excite state ($J^\pi_k$)  is given by the reduced matrix element of the ISM operator as
\begin{eqnarray}
&&B(IS0;\textrm{g.s.}\to J^\pi_k)\nonumber\\
&&=\frac{1}{2J+1}
|\langle \textrm{g.s.}||{\cal M}(IS0)|| J^\pi_k\ \rangle |^2.
\end{eqnarray}
The energy-weighted sum (EWS) of the ISM strengths is defined 
as
\begin{eqnarray}
S(IS0)\equiv \sum_{k} (E_{k}-E_\textrm{g.s.}) 
B(IS0;\textrm{g.s.}\to J^\pi_k),
\end{eqnarray}
If the interaction commutes with ${\cal M}({\rm IS0})$, 
the ISM energy weighted sum rule (EWSR) 
\begin{eqnarray}
S(IS0) =\frac{2\hbar^2}{m}A\langle r^2 \rangle_\textrm{g.s.}
 \label{eq:eqsr}
\end{eqnarray}
is satisfied. Here $\langle r^2 \rangle_\textrm{g.s.}$ is the mean-square radius of the ground state and equals to 
$\langle \textrm{g.s.}|{\cal M}(IS0) |\textrm{g.s.}\rangle/A$.

\section{Results} \label{sec:results}
\subsection{Effective nuclear forces} \label{subsec:effectiveforce}
The effective Hamiltonian consists of the single-particle kinetic terms $t_i$ and two-body forces
$v_{ij}$ containing effective nuclear forces and the Coulomb force, 
\begin{equation}
H_{\rm eff}=\sum_i t_i -T_G+\sum_{i<j} v_{ij},
\end{equation}
where the kinetic energy $T_G$ of the center of mass motion is subtracted. 
As for the effective nuclear forces,  
the Volkov No.2 force ~\cite{Volkov} is used for the central force, 
and the spin-orbit term of the G3RS force~\cite{LS} is adopted for the
spin-orbit force, as done in preceding works for 
$^{10}$Be and $^9$Li structures \cite{Dote:1997zz,Itagaki:1999vm,Itagaki:2000nn,Suhara:2009jb,KanadaEn'yo:2011nc}.
The interaction parameters are $(b=h=0.125, m=0.60)$ for the Volkov No.2 force and
$u_{I}=-u_{II}=1600$ MeV for the strength of the spin-orbit force. 
These are the same parameters as those used in
Refs.~\cite{Suhara:2009jb,KanadaEn'yo:2011nc}.
The Coulomb force is approximated by seven-range Gaussians.

\subsection{Energy levels of $^9$Li and $^{10}$Be}

As described in \eqref{eq:gcm}, 
the $\He+\alpha(t)$ cluster wave functions $\ket{\Phi_{\tau}(D)}$
specified by the label $\tau$ and the distance parameter $D$ are superposed. 
In the default GCM calculation, I take the generator coordinate, $D=1,2,\cdots, 8$ fm, 
which corresponds to a bound state approximation.
To see resonance features, I also take a larger model space, $D=1,2,\cdots, 15$ fm,
to examine the effect of coupling with discretized continuum states.
The six configurations, $\tau$=\{a,b,c,d,e,f\}, are adopted at each $D$, and totally 
$6\times 8=48 (6\times 15=90)$ basis wave functions are superposed in the $D=1,2,\cdots, 8$ fm ($D=1,2,\cdots, 15$ fm) calculation, 
which I denote the ``$D\le 8$ ($D\le 15$)'' calculation. In the $D\le 8$ and 
$D\le 15$ calculations, 
the $K$-mixing is taken into account. 
In addition, I also perform the GCM calculation with a truncated model space 
using only the transverse configurations, (c) and (f), denoted the ``(c+f)'' calculation.
In the (c+f) calculation, the angular momentum projected $\He+\alpha(t)$ wave functions of the configurations, (c) and (f), 
with $K=0(1/2)$ and $D=1,2,\cdots, 8$ fm are used.

\begin{figure}[th]
\begin{center}
\includegraphics[width=7.5cm]{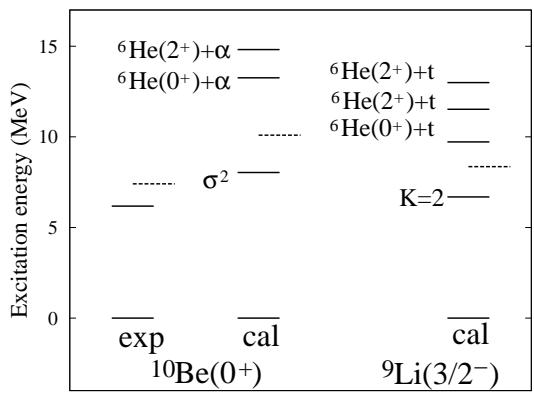} 	
\end{center}
\caption{Energy levels of $\Be(0^+)$ and $\Li(3/2^-)$ 
obtained by the $D\le 8$ calculation. 
The experimental energy levels of $\Be(0^+)$ are also shown.
The $\He+\alpha$ and $\He+t$ threshold energies are shown by dashed lines.}
\label{fig:spe}
\end{figure}

Figure \ref{fig:spe} shows 
the energy spectra of $\Be(0^+)$ and $\Li(3/2^-)$ obtained
by the default $D\le 8$ calculation. The $\He+\alpha$ and $\He+t$ threshold energies are shown by
dashed lines.
The $\Be(0^+_{3,4})$ and $\Li(3/2^-_{3,4,5})$ states are obtained
above the $\He+\alpha$ and $\He+t$ threshold energies, respectively.
The $\Be(0^+_3)$ and $\Be(0^+_4)$ states 
have relatively larger $\He(0^+)+\alpha$ and  $\He(2^+)+\alpha$ components, respectively.
The $\Li(3/2^+_3)$ is dominated by the $\He(0^+)+t$ component, 
where as the $\Li(3/2^+_4)$ and $\Li(3/2^+_5)$ 
have the dominant $\He(2^+)+t$ components. 
The dominant $\He+\alpha(t)$ 
components decrease at the boundary, $D=8$ fm, and hence these states are 
regarded as resonance states. 
Other states higher than the $\Be(0^+_4)$ and $\Li(3/2^+_5)$ states 
do not show such the resonance feature and are regarded as continuum states.
In the low-energy region of $\Be$, 
the $0^+_2$ state of $\Be$ with the molecular $\sigma$-orbital structure is obtained
below the threshold energy.
This state is the band-head state of the $K^\pi=0^+_2$ band and 
assigned to the experimental $0^+$ state at 6.18 MeV. 
However, in $\Li$,  the molecular orbital structure does not appear because the $\sigma$-orbital is 
unfavored around the $\alpha+t$ core because of the asymmetry of the core potential 
as discussed in the previous work. 
The $\Li(3/2^-_2)$ state obtained below the $\He+t$ threshold energy
is the band-head state of the $K^\pi=3/2^-$ band.

In the $D\le 8$ calculation,
the resonance states are obtained as bound state solutions in 
the model space of  $D \le 8$ fm.
I estimate the partial decay widths $\Gamma_{I\otimes J'}$
of the resonance states for $\He(I^+)+\alpha(t)$ channels 
with the angular momentum coupling $[I\otimes J']_{J}$
from the reduced width amplitude
$y(a)$ at a channel radius $a$
as described in appendix \ref{app:width}.
Here $J'$ is the resultant angular momentum of  the angular momenta
$I'$ and $l$, where $I'$ is the internal angular momentum of 
the $\alpha$($t$) cluster and $l$ is the orbital angular momentum
for the relative coordinate $\bvec{r}$ between centers of 
mass of two clusters.  
For $\He(0^+)+\alpha$ and $\He(2^+)+\alpha$ decays  
of $\Be(0^+)$, $S$- and $D$-wave decays in the $[I\otimes J']_{J}=  [0\otimes 0]_{0}$ and 
$[2\otimes 2]_{0}$ channels are calculated, respectively. 
For $\He(0^+)+t$ and $\He(2^+)+t$ decays of $\Li(3/2^-)$, I consider 
$P$-wave decays in three channels, $[0\otimes 3/2]_{3/2}$, 
$[2\otimes 3/2]_{3/2}$, and $[2\otimes 1/2]_{3/2}$.

The calculated partial decay widths $\Gamma_{I\otimes J'}$
at channel radii $a=$5, 6, and 7 fm are shown in Table \ref{tab:gamma}. 
Here, decay energies ($E_{\rm decay}$) are calculated from the theoretical energies of $t$, $\alpha$, $\He(0^+,2^+)$, $\Be$, and $\Li$. 
The sum  $\Gamma_{\rm sum}$ of the partial widths and the dimensionless reduced widths $\theta^2(a)=(a/3)|ay(a)|^2$ are also shown in the table.
Here after, I discuss the decay widths 
calculated at the channel radius that gives the largest $\Gamma_{\rm sum}$ for each state. 
The calculated width of the $\Be(0^+_3)$ is 
$\Gamma_{\rm sum}=1.6$ MeV with the dominant 
$\He(0^+)+\alpha$ decay and that of the $\Be(0^+_4)$ is
 $\Gamma_{\rm sum}=1.0$ MeV with comparable
partial widths of $\He(0^+)+\alpha$ and $\He(2^+)+\alpha$ decays.
The calculated widths of the 
$\Li(3/2^-_3)$, $\Li(3/2^-_4)$, 
and $\Li(3/2^-_5)$ are 
$\Gamma_{\rm sum}=0.65$, 0.75, and 2.0 MeV,
respectively. The $\Li(3/2^-_3)$ can decay only
in the $\He(0^+)+t$ channel because the $\He(2^+)+t$ channel is closed. The $\Li(3/2^-_4)$ has
comparable partial widths of $\He(0^+)+t$ and $\He(2^+)+t$ decays, whereas
the $\Li(3/2^-_5)$ has dominant $\He(2^+)+t$ decays. 

\begin{figure}[th]
\begin{center}
\includegraphics[width=0.48\textwidth]{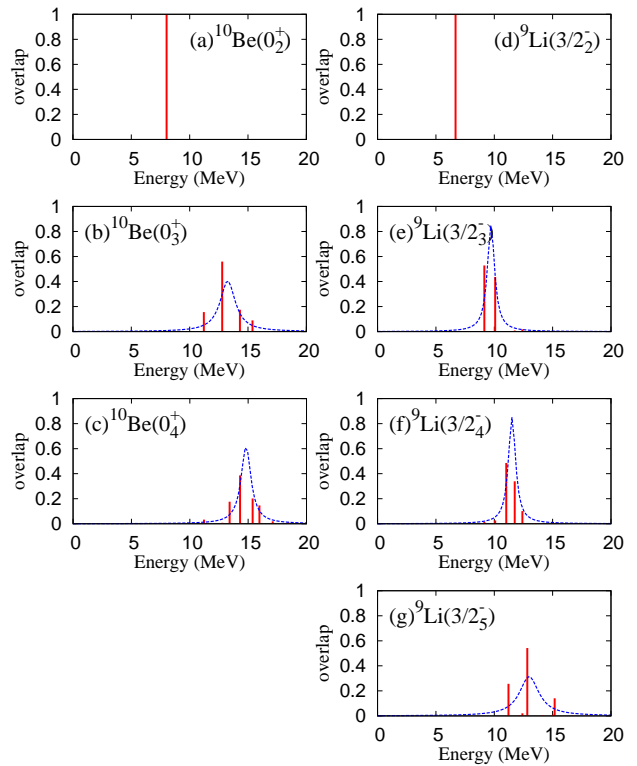} 	
\end{center}
\caption{Distributions of 
the squared overlap $|\langle \Psi^{J^\pi_k}_{D\le 8}|\Psi^{J^\pi_l}_{D\le 15}\rangle|^2$ of the $J_k$ state obtained by the $D\le 8$ calculation 
with states obtained by the $D\le 15$ calculation.
Distributions of the $J_k=0^+_2$, $0^+_3$, and $0^+_4$ states in $\Be$
are shown in (a)-(c), and those of  
the $J_k=3/2^-_2$, $3/2^-_3$, $3/2^-_4$, and $3/2^-_5$ states in $\Li$
are shown in (d)-(g). 
The Breit-Wigner distributions at the energies ($E_x$) with 
the widths ($\Gamma_{\rm sum}/2$) for the resonances obtained by the $D\le 8$
calculation are shown by dashed lines. The largest value of 
$\Gamma_{\rm sum}$ at $a=5$, 6, 7 fm table \ref{tab:gamma} is adopted for each state.
}
\label{fig:dia48in90}
\end{figure}

To see resonance features of these states obtained above the threshold energies, 
I perform the $D\le 15$ calculation with 
the larger model space than that of the $D\le 8$ calculation. 
In the model space enlarged from $D\le 8$ fm to $D\le 15$ fm,
the resonance states $\ket{\Psi^{J^\pi_k}_{D\le 8}}$, 
which obtained as bound state solutions in the $D\le 8$ calculation, 
couple with discretized continuum states and their components 
are fragmented in states ($\ket{\Psi^{J^\pi_l}_{D\le 15}}$) in the $D\le 15$ calculation.
In Fig.~\ref{fig:dia48in90}, 
I show the squared overlap 
$|\langle \Psi^{J^\pi_k}_{D\le 8}|\Psi^{J^\pi_l}_{D\le 15}\rangle|^2$.
The overlap distributions show that 
the components of the resonance states 
obtained by the $D\le 8$ calculation are fragmented into several states 
in the $D\le 15$ calculation, whereas those  
of the $\Be(0^+_2)$ and $\Li(3/2^+_2)$ below 
the threshold energies are not fragmented. 
As shown in the figure, the overlap distributions are consistent with   
the Breit-Wigner distributions at the resonance energies with 
the widths ($\Gamma_{\rm sum}/2$) obtained by the $D\le 8$
calculation. This result indicates that the decay widths $\Gamma_{\rm sum}$
estimated by using the reduced width amplitudes are reasonable. 

$\He+\alpha$ cluster resonances as well as molecular orbital states in $\Be$ have 
been investigated by Ito {\it et al.} with the GTCM.
The GTCM calculation predicted a $\He(2^+)+\alpha$ resonance  
at the energy $E_r=3.6$ MeV relative to the $\He+\alpha$  threshold energy and 
a $\He(0^+)+\alpha$ state as a broad continuum state in the $E_r=1-4$ region. 
The present result of the $\Be(0^+_4)$ with the dominant
$\He(2^+)+\alpha$ component may correspond to the $\He(2^+)+\alpha$ state of
the GTCM. The $\Be(0^+_3)$ obtained in the present result, which 
has the dominant $\He(0^+)+\alpha$ component 
and a larger width than the $\Be(0^+_4)$, is likely to 
correspond to the  $\He(0^+)+\alpha$ state
of the GTCM.

\begin{table}[htb]
\caption{
\label{tab:gamma} 
Partial decay widths $\Gamma_{I\otimes J'}$ (MeV)
for $\He(I^+)+\alpha$ of $\Be(0^+)$ and $\He(I^+)+t$ of $\Li(3/2^-)$
obtained by the $D\le 8$ calculation. $\Gamma_{I\otimes J'}$
for $S$-wave and $D$-wave decays
in $[0\otimes 0]_0$ and $[2\otimes 2]_0$ of $\Be(0^+)$ and 
$P$-wave decays in $[0\otimes 3/2]_{3/2}$, $[2\otimes 3/2]_{3/2}$, and $[2\otimes 1/2]_{3/2}$
of $\Li(3/2^-)$ are shown. 
The sum ($\Gamma_{\rm sum}$) of the partial widths and the dimensionless reduced widths $\theta^2(a)=(a/3)|ay(a)|^2$
are also shown. 
}
\begin{center}
\begin{tabular}{cccccccc}
\hline
	&	$E_{\rm decay}$ (MeV)	&	\multicolumn{3}{c}{$\theta^2$}&	\multicolumn{3}{c}{$\Gamma_{I\otimes J'}$ (MeV)}	\\
$a$ (fm) &		&	5	&	6	&	7	&	5	&	6	&	7 \\
 \multicolumn{8}{l}{$\He(0^+)+\alpha$ with $[0\otimes 0]_0$}  \\
$\Be(0^+_3)$	&	3.2 	&	0.04 	&	0.20 	&	0.38 	&	0.18 	&	0.92 	&	1.55 	\\
$\Be(0^+_4)$	&	4.7 	&	0.02 	&	0.01 	&	0.10 	&	0.14 	&	0.05 	&	0.54 	\\			
 \multicolumn{8}{l}{$\He(2^+)+\alpha$ with $[2\otimes 2]_0$}  \\
$\Be(0^+_3)$	&	1.3 	&	0.13 	&	0.12 	&	0.08 	&	0.011 	&	0.02 	&	0.02 	\\
$\Be(0^+_4)$	&	2.9 	&	0.07 	&	0.19 	&	0.23 	&	0.10 	&	0.34 	&	0.49 	\\		
 \multicolumn{8}{l}{$\He(0^+)+t$ with $[0\otimes 3/2]_{3/2}$}  \\
$\Li(3/2^-_3)$	&	1.4 	&	0.26 	&	0.33 	&	0.29 	&	0.59 	&	0.75 	&	0.65 	\\
$\Li(3/2^-_4)$	&	3.2 	&	0.01 	&	0.04 	&	0.06 	&	0.08 	&	0.21 	&	0.30 	\\
$\Li(3/2^-_5)$	&	4.6 	&	0.000 	&	0.001 	&	0.004 	&	0.000 	&	0.007 	&	0.02 	\\												
 \multicolumn{8}{l}{$\He(2^+)+t$ with $[2\otimes 3/2]_{3/2}$}  \\
$\Li(3/2^-_4)$	&	1.3 	&	0.13 	&	0.15 	&	0.13 	&	0.30 	&	0.33 	&	0.27 	\\
$\Li(3/2^-_5)$	&	2.8 	&	0.06 	&	0.15 	&	0.20 	&	0.33 	&	0.69 	&	0.86 	\\											
 \multicolumn{8}{l}{$\He(0^+)+t$ with $[2\otimes 1/2]_{3/2}$}  \\
$\Li(3/2^-_4)$	&	1.3 	&	0.09 	&	0.10 	&	0.08 	&	0.19 	&	0.21 	&	0.18 	\\
$\Li(3/2^-_5)$	&	2.8 	&	0.09 	&	0.20 	&	0.27 	&	0.44 	&	0.93 	&	1.15 	\\								
&	$E_r$ (MeV)	&	\multicolumn{3}{c}{$\Gamma_{\rm sum}$} (MeV)	&&&	\\
$\Be(0^+_3)$	&	3.2 &	0.19 	&	0.93 	&	1.6 	&		&		&		\\
$\Be(0^+_4)$	& 4.7		&	0.24 	&	0.39 	&	1.0 	&		&		&		\\
$\Li(3/2^-_3)$	&	1.4 	&	0.59 	&	0.75 	&	0.65 	&		&		&		\\
$\Li(3/2^-_4)$	& 3.2 	&	0.57 	&	0.75 	&	0.75 	&		&		&		\\
$\Li(3/2^-_5)$	&	4.6 	&	0.77 	&	1.6 	&	2.0 	&		&		&		\\
\hline	
\end{tabular}
\end{center}
\end{table}


\subsection{Isoscalar monopole transition strengths}
\begin{figure}[th]
\begin{center}
\includegraphics[width=7.5cm]{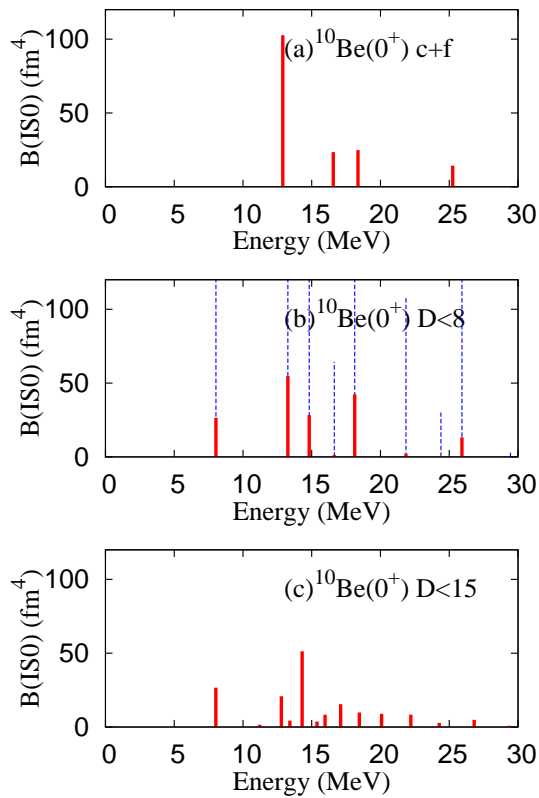} 	
\end{center}
\caption{\label{fig:be10-is0}
ISM strength distributions
of $\Be(0^+)$ 
obtained by the (c+f) , $D\le 8$, and  $D\le 15$ calculations. 
Dashed lines in the middle panel (b) for the $D\le 8$ calculation 
show fiftyfold values of the strengths.}
\end{figure}

\begin{figure}[thb]
\begin{center}
\includegraphics[width=7.5cm]{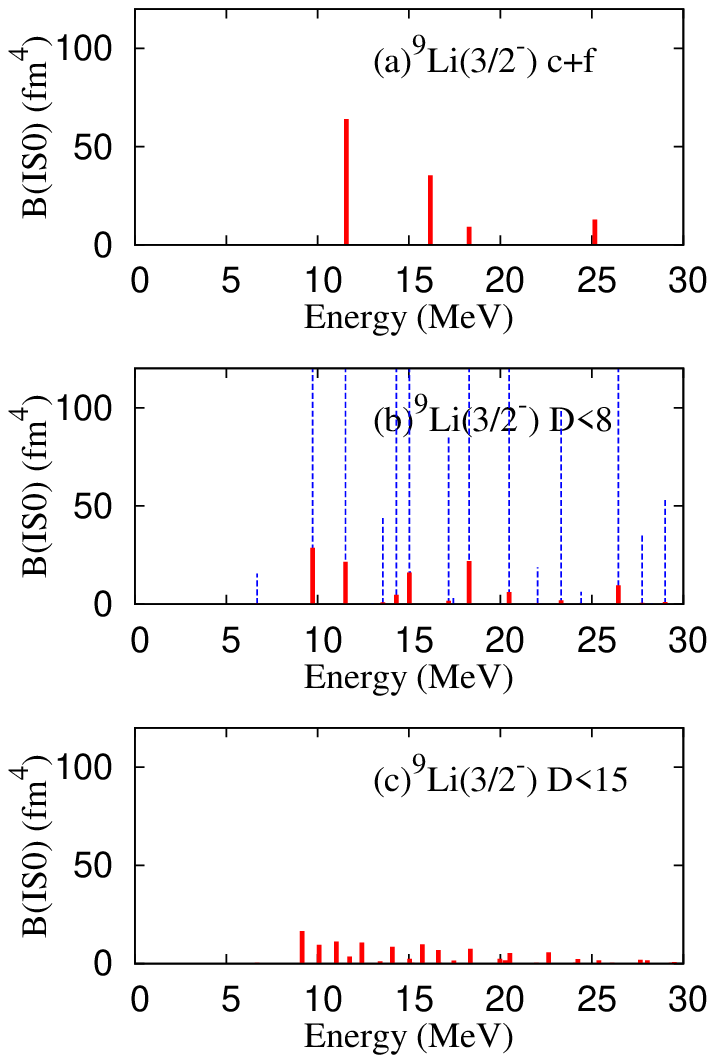} 	
\end{center}
\caption{\label{fig:li9-is0}
ISM strength distributions
of $\Li(3/2^-)$ 
obtained by the (c+f) , $D\le 8$, and  $D\le 15$ calculations. 
Dashed lines in the middle panel (b) for the $D\le 8$ calculation 
show fiftyfold values of the strengths.
}
\end{figure}

\begin{figure}[thb]
\begin{center}
\includegraphics[width=0.48\textwidth]{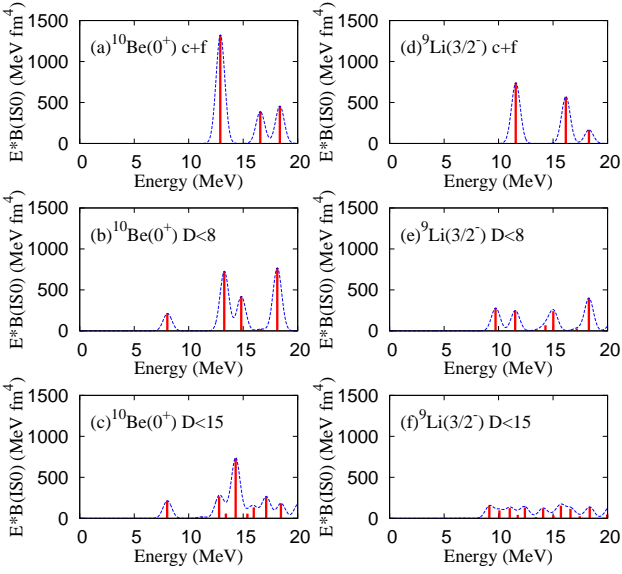} 	
\end{center}
\caption{\label{fig:ewis0}
Energy weighted ISM strengths (solid lines).
(a)-(c) The distributions in $\Be(0^+)$ obtained by the (c+f), $D\le 8$, $D\le 15$ calculations, and 
(d)-(f) the distributions in $\Li(3/2^-)$.
The Gaussian smeared distributions of the energy weighted ISM strengths with the fixed width 
$1/\sqrt{\pi}$ are shown by dashed lines. 
}

\end{figure}

The ISM transition strengths 
of $\Be(0^+)$ and $\Li(3/2^-)$ 
calculated by the (c+f), $D\le 8$, and $D\le 15$ calculations
are shown in Figs.~\ref{fig:be10-is0} and \ref{fig:li9-is0}.
The energy weighted ISM strengths and their Gaussian smeared distributions
are shown in Fig.~\ref{fig:ewis0}. 
The proton and matter radii,  
EWSR,  EWS, and the ratio EWS/EWSR obtained by three calculations 
are listed in table \ref{tab:is0}.
The experimental radii are also shown in the table.
In the (c+f) calculation, the model space is truncated and contains 
only the configurations (c) and (f), which corresponds to the 
transverse configurations of $\He$ with two neutrons in $p_x$ or $p_y$ orbits.
In spite of the truncation of $\He$ configurations,
the EWS of the (c+f) result is consistent with those of 
the $D\le 8$ and $D\le 15$ calculations. The reason is that the
ground state $|\textrm{g.s.}\rangle$ and also its ${\cal M}(IS0)$ operated state
${\cal M}(IS0)|\textrm{g.s.}\rangle$ are contained in 
the truncated model space of the transverse configurations (c) and (f). 

In the (c+f) result of $\Be$, 
the ISM strength is concentrated on the first excited state
at $E_x \sim 13$ MeV. In the $D\le 8$ result, the ISM strengths are split
by coupling with other configurations (a), (b), (d), (e). 
However, the significant strengths remain for the transitions to 
$\Be(0^+_3)$ and $\Be(0^+_4)$ states in the $E_x= 13-15$ MeV. 
In the $D\le 15$ result, the strengths are fragmented further because of the 
coupling with the continuum states, but the strengths
are still concentrated in the $E_x= 13-15$ MeV region. 
In the energy weighted strength
distributions shown in Fig.~\ref{fig:ewis0}(c) ,  
the enhancement of the ISM strengths are found in this energy region.
This result indicates that
the ISM excitation can be a good probe to observe the $\He+\alpha$ cluster resonances
in $\Be$. The ISM strength for the $\Be(0^+_2)$ is not so remarkable compared with those in the $E_x= 13-15$ MeV region. 

In the (c+f) result of $\Li$, 
the ISM strength is somewhat concentrated on the first excited state
at $E_x \sim 11$ MeV, but the magnitude of the strength 
is not so remarkable as the case of $\Be$.
The strength distributions are  
fragmented in the $D\le 8$ result because of the mixing 
of other configurations (a), (b), (d), (e), 
and those are strongly scattered in the $D\le 15$ result
because of the coupling with continuum states. 
The stronger fragmentation of the ISM strengths in $\Li$ than 
that in $\Be$ originates in a variety of
angular momentum channels $[I\times J']$ 
and $K$-mixing in the total spin-parity $J^\pi=3/2^-$ of final states.
Consequently, 
there is no concentration of the ISM strengths on
the $\He+t$ cluster resonances in $\Li$.
Indeed, as shown in Fig.~\ref{fig:ewis0}(e), 
the energy weighted strengths are widely distributed and no remarkable 
strengths to the resonance states in $\Li$.
The $\Li(3/2^-_2)$ has almost no ISM strength, because this 
state is the band-head state of the $K^\pi=3/2^-$ band and is not
excited by the ISM operator from the $\Li$ ground state, which
has the dominant $K^\pi=1/2^-$ component.

\begin{table}[htb]
\caption{
\label{tab:is0} 
Theoretical values of proton and matter radii (fm),  and 
the EWSR (MeV fm$^4$), EWS (MeV fm$^4$), 
and the ratio EWS/EWSR of the ISM strengths 
obtained by the $D\le 8$ calculation. 
The experimental proton radii reduced from the charge radii \cite{Nortershauser:2008vp,Ewald:2004zz}, and 
the experimental matter radii deduced from 
the interaction cross sections \cite{ozawa2001} 
are also shown.
}
\begin{center}
\begin{tabular}{cccccc}
\hline
	\multicolumn{6}{c}{$\Be$}	\\
	&	$r_p$&		$r_m $	&	EWSR&	EWS&	EWS/EWSR \\	
(c+f)	&	2.31 	&	2.33 	&	4.5$\times 10^2$ 	&	2.7$\times 10^2$ 	&	0.60 	\\
$D\le 8$	&	2.31 	&	2.34 	&	4.5$\times 10^2$ 	&	2.7$\times 10^2$ 	&	0.61 	\\
$D\le 15$	&	2.31 	&	2.34 	&	4.5$\times 10^2$ 	&	2.7$\times 10^2$ 	&	0.61 	\\
exp.	&	2.221(18)	&	2.30(2)	&		&		&		\\
		\multicolumn{6}{c}{$\Li$}	\\								
(c+f)&	2.12 	&	2.23 	&	3.7$\times 10^2$ 	&	1.68$\times 10^2$ 	&	0.45 	\\
$D\le 8$	&	2.11 	&	2.22 	&	3.7$\times 10^2$ 	 	&	1.68$\times 10^2$ 	&	0.46 	\\
$D\le 15$	&	2.11 	&	2.22 	&	3.7$\times 10^2$ 	 	&	1.68$\times 10^2$ 	&	0.46 	\\
exp.		&	2.05(4)	&	2.32(2)	&		&		&		\\
\hline	
\end{tabular}
\end{center}
\end{table}


\section{Discussions} \label{sec:discussions}
In this section, I discuss cluster structures of $\Be(0^+)$ and $\Li(3/2^-)$ 
in connection with ISM excitations.
To analyze cluster structures, I evaluate cluster components at a certain distance 
by calculating overlaps of the 
obtained $\Be(0^+)$ and $\Li(3/2^-)$ wave functions with 
$\He+\alpha(t)$ wave functions specified by the distance parameter $D$.

\subsection{Reference $\He(I^+)+\alpha(t)$ wave functions with $D$} 

I consider two types of reference wave functions at a certain distance $D$. 
One is the $\He(I^+)+\alpha(t)$ wave functions 
with the angular momentum coupling $[I\otimes J']_J$, 
and the other is the basis wave functions $\Phi_\tau(D)$ used in the GCM 
calculation. 
The former is the weak-coupling $\He+\alpha(t)$ wave functions, in which
the internal angular momenta of clusters and the orbital angular momentum of the inter-cluster motion are weakly coupled, whereas the latter is the strong-coupling 
$\He+\alpha(t)$ wave functions, in which a deformed $\He$ cluster is oriented 
to a specific angle from the $\alpha(t)$ direction. 

The $\He(I^+)+\alpha(t)$ wave functions 
are constructed by using the ground and the first excite states 
$\He(0^+)$ and $\He(2^+)$ of 
an isolated $\He$ cluster described by ho 
$p$-shell configurations. The $\He(I^+)+\alpha(t)$ wave functions with the distance $D$
are given by linear combination of the basis wave functions 
$P^{J\pi}_{MK} \ket{\Phi_{\tau}(D)}$ used in the present model, 
and they are defined as,
\begin{eqnarray}
&&\Phi^{[I\otimes J']_J}_{\He+\alpha(t)}(D)=\nonumber \\
&&n_0{\cal A}\left\{\phi_{\rm G}(\bvec{R}) \gamma_l(D;r)
\left[\varphi^{\He}_{I}\left[\varphi^{\alpha(t)}_{I'}Y_l(\hat r)\right]_{J'}\right]_J 
\right\}, \\
&& \gamma_l(D;r) \equiv 4\pi (\frac{2\tilde\nu}{\pi})^{\frac{3}{4}} i_l(2\tilde\nu D r) e^{-\tilde\nu(r^2+D^2)},\\
&& \tilde \nu \equiv  \frac{A_1A_2}{A} \nu,\\
&&\phi_{\rm G}(\bvec{R})=\left( \frac{2A\nu}{\pi} \right) e^{-A\nu\bvec{R}^2},
\end{eqnarray}
where $i_l$ is the modified spherical Bessel function,
$\varphi^{\He}_{I}$ and $\varphi^{\alpha(t)}_{I'}$ are the internal wave functions of
$\He(I^+)$ and $\alpha(t)$ clusters with the internal angular momentum 
$I$ and $I'$, respectively.  
$\phi_{\rm G}$ is the wave function of the center of mass motion and
$n_0$ is the normalization factor.

For the strong-coupling 
$\He+\alpha(t)$ wave functions, I consider the
$J^\pi$ and $K$ projected states of the basis wave functions $\Phi_\tau(D)$ 
with specific configurations as, 
\begin{eqnarray}
&&\Phi^{(T)}_{\He+\alpha(t)}(D)\equiv n_0P^{J\pi}_{M0(1/2)} \ket{\Phi_{\tau=f}(D)}, \\
&&\Phi^{(A)}_{\He+\alpha(t)}(D)\equiv n_0P^{J\pi}_{M0(1/2)} \ket{\Phi_{\tau=a}(D)}, \\
&&\Phi^{(I_z2)}_{\He+t}(D)\equiv n_0P^{J\pi}_{M,-3/2} \ket{\Phi_{\tau=c}(D)}.
\end{eqnarray}
$\Phi^{(T)}_{\He+\alpha(t)}(D)$ is the $jj$ coupling  transverse (T) configuration 
corresponding to the configuration (f) with two neutrons
in the transverse orbits with $|j_z|=3/2$, 
whereas  $\Phi^{(A)}_{\He+\alpha(t)}(D)$ is the $ls$ coupling 
aligned (A) configuration given by the configuration (a) 
with two neutrons coupling to $S=0$ in the aligned orbit $p_z$.
In a short distance ($D$) region, configurations
other than transverse configurations feel strong Pauli blocking, and therefore, 
$\Phi^{(T)}_{\He+\alpha(t)}(D)$ is most favored.
Indeed, the ground state wave function has the largest overlap with  $\Phi^{(T)}_{\He+\alpha(t)}(D)$ with $D=2-3$ fm as shown later. 
The configuration (a), i.e., $\Phi^{(A)}_{\He+\alpha(t)}(D)$ corresponds to the 
molecular $\sigma$-orbital structure, and
this component with $D=4-5$ fm has a large overlap with the $\Be(0^+_2)$.
$\Phi^{(I_z2)}_{\He+t}(D)$ is the $K=-3/2$ state projected from 
the $ls$ coupling transverse configuration given by the configuration (c), 
in which a $S=0$ two-neutron pair 
in the $\He$ cluster is rotating to give $I_z=-2$. $\Phi^{(I_z2)}_{\He+t}(D)$ 
with $D=2-3$ fm is the 
dominant component of the $\Li(3/2^-_2$).  

\begin{figure}[th]
\begin{center}
\includegraphics[width=7cm]{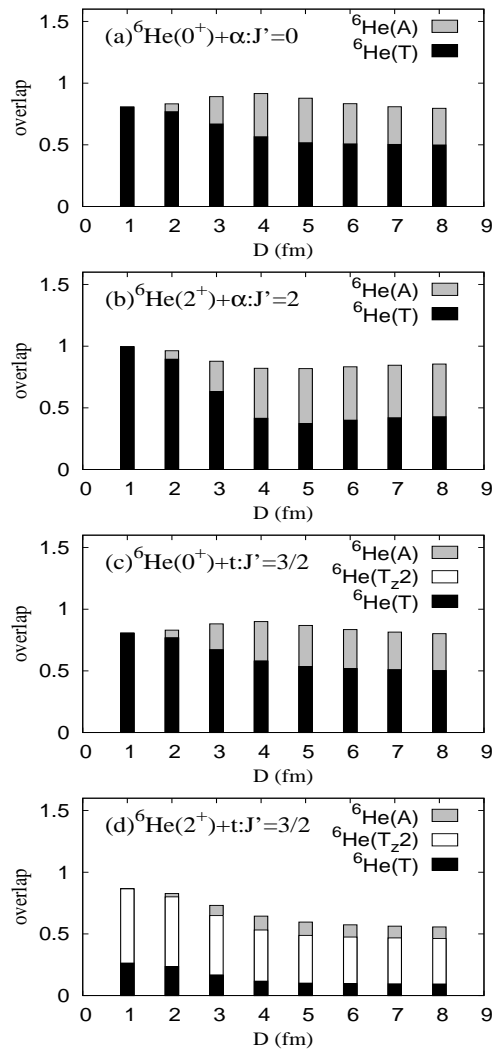} 	
\end{center}
\caption{
Squared overlaps between the weak-coupling cluster wave functions 
$\Phi^{[I\otimes J']_J}_{\He+\alpha(t)}(D)$ and the 
strong-coupling cluster wave functions $\Phi^{(T,A,I_z2)}_{\He+\alpha(t)}(D)$.
The squared overlaps of 
(a) $\Phi^{[0\otimes 0]_0}_{\He+\alpha}(D)$, (b) $\Phi^{[2\otimes 2]_0}_{\He+\alpha}(D)$, (c) $\Phi^{[0\otimes 3/2]_{3/2}}_{\He+t}(D)$, and (d) 
$\Phi^{[0\otimes 3/2]_{3/2}}_{\He+t}(D)$ with $\Phi^{(T,A,I_z2)}_{\He+\alpha(t)}(D)$ 
are shown.
}
\label{fig:b1-base}
\end{figure}

Thus defined weak-coupling and strong-coupling reference wave functions, 
$\Phi^{[I\otimes J']_J}_{\He+\alpha(t)}(D)$ and $\Phi^{(T,A,I_z2)}_{\He+\alpha(t)}(D)$, 
are not orthogonal to each other. 
In Fig.~\ref{fig:b1-base},  I show 
squared overlaps of 
$\Phi^{[I\otimes J']_J}_{\He+\alpha(t)}(D)$ with $\Phi^{(T,A,I_z2)}_{\He+\alpha(t)}(D)$.
In a short distance region, only transverse configurations are Pauli allowed
but other configurations feel strong Pauli blocking
because of the antisymmetrization effect 
between clusters. As a result, both of the $\He(0^+)+\alpha$ and $\He(2^+)+\alpha$ 
wave functions, 
$\Phi^{[0\otimes 0]_0}_{\He+\alpha}(D)$ and 
$\Phi^{[2\otimes 2]_0}_{\He+\alpha}(D)$, have dominant overlaps with 
the transverse configuration $\Phi^{(T)}_{\He+\alpha}(D)$
but no overlap with the aligned configuration $\Phi^{(A)}_{\He+\alpha}(D)$.
This means that the 
$\He(0^+)+\alpha$ and $\He(2^+)+\alpha$ wave functions at the short distance 
are almost equivalent to the $\Phi^{(T)}_{\He+\alpha}(D)$ 
and contain no $\Phi^{(A)}_{\He+\alpha}(D)$ component. 
As for the $\He+t$ wave functions at a short distance $D$, 
the $\He(0^+)+t$ wave function is almost equivalent to
 $\Phi^{(T)}_{\He+t}(D)$, 
the $\He(2^+)+t$ wave function is a mixing of $\Phi^{(T)}_{\He+t}(D)$
and $\Phi^{(I_z3/2)}_{\He+t}(D)$, whereas they
contain no $\Phi^{(A)}_{\He+t}(D)$ component.
This is a trivial consequence of the antisymmetrization of  the $\He+\alpha$ cluster wave functions, 
and it indicates that,  at a short distance, the weak-coupling $\He+\alpha(t)$ wave functions have less physical meaning  than the strong-coupling 
$\He+\alpha(t)$ wave functions.
In an enough large distance $D$ region free from the 
Pauli blocking between two clusters, $\Phi^{[I\otimes J']_J}_{\He+\alpha(t)}(D)$
with different $I$ and $J'$ is orthogonal to each other, and has overlaps with 
$\Phi^{(T,A,I_z2)}_{\He+\alpha(t)}(D)$ in specific ratios.

\subsection{Cluster structures of $\Be(0^+)$ and $\Li(3/2^-)$}
\begin{figure}[th]
\begin{center}
\includegraphics[width=0.48\textwidth]{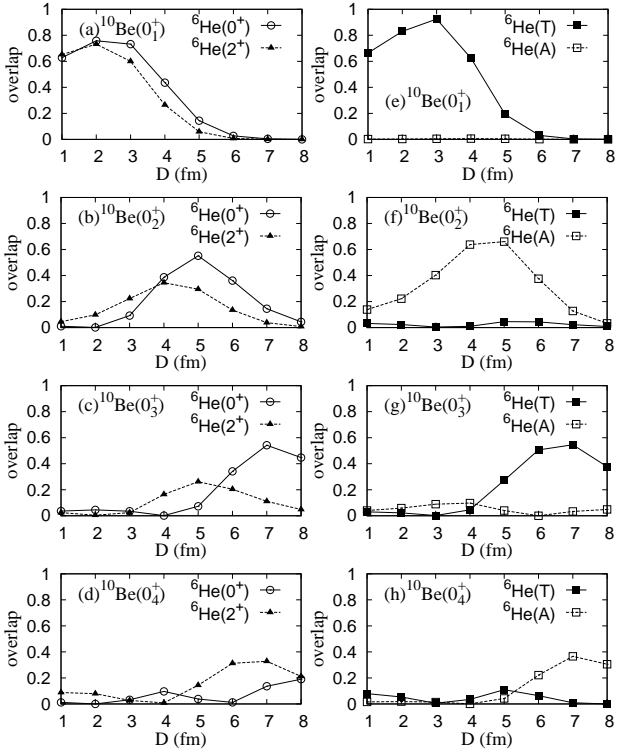} 	
\end{center}
\caption{$\He+\alpha$ components in 
$\Be(0^+)$  obtained by the $D\le 8$ calculation. 
(a-d) Squared overlaps of $\Be(0^+)$ 
with $\Phi^{[0\otimes 0]_0}_{\He+\alpha}(D)$ and 
$\Phi^{[2\otimes 2]_0}_{\He+\alpha}(D)$.  
(e-h) Squared overlaps of $\Be(0^+)$  
with  $\Phi^{(T)}_{\He+\alpha}(D)$ and $\Phi^{(A)}_{\He+\alpha}(D)$.
\label{fig:be10-d1-8}}
\end{figure}

\begin{figure}[th]
\begin{center}
\includegraphics[width=0.48\textwidth]{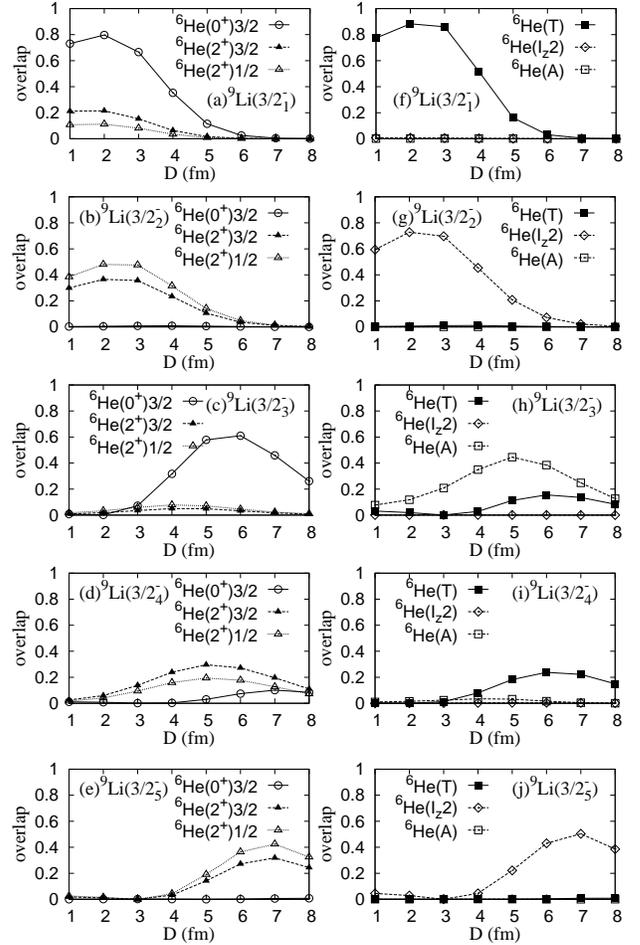} 	
\end{center}
\caption{$\He+t$  components in 
$\Li(3/2^-)$ obtained by the $D\le 8$ calculation. 
(a-e) Squared overlaps of $\Li(3/2^-)$ 
with 
$\Phi^{[0\otimes 3/2]_{3/2}}_{\He+t}(D)$, 
$\Phi^{[2\otimes 3/2]_{3/2}}_{\He+t}(D)$, and 
$\Phi^{[2\otimes 1/2]_{3/2}}_{\He+t}(D)$.
(f-j) Squared overlaps of the $\Li(3/2^-)$  
with 
 $\Phi^{(T)}_{\He+t}(D)$, $\Phi^{(A)}_{\He+t}(D)$,  and $\Phi^{(I_z2)}_{\He+t}(D)$.
\label{fig:li9-d1-8}
}
\end{figure}

To analyze cluster structures of the 
obtained $\Be(0^+)$ and $\Li(3/2^-)$, I calculate the overlaps with 
the reference $\He+\alpha(t)$ wave functions at the distance $D$, 
$\Phi^{[I\otimes J']_J}_{\He+\alpha(t)}(D)$ and $\Phi^{(T,A,I_z2)}_{\He+\alpha(t)}(D)$.
Figures \ref{fig:be10-d1-8} and \ref{fig:li9-d1-8} show the 
squared overlaps of the  $\Be(0^+)$ and $\Li(3/2^-)$ obtained by the $D\le 8$ calculation 
with the $\He+\alpha(t)$ wave functions  plotted as functions of $D$.


The $\Be$ ground state has the dominant overlap 
with the transverse configuration
$\Phi^{(T)}_{\He+\alpha}(D)$ at $D\sim 3$ fm and almost no overlap with 
the aligned configuration $\Phi^{(A)}_{\He+\alpha}(D)$. 
The $\Be(0^+_2)$ is dominated by $\Phi^{(A)}_{\He+\alpha}(D)$
at $D=4-5$ fm. This is consistent with the molecular 
$\sigma$-orbital structure of the $\Be(0^+_2)$ as discussed in the preceding works.
Thus, the low-lying states, $\Be(0^+_1)$ and  $\Be(0^+_2)$, can 
be clearly classified by the strong-coupling
$\He+\alpha$ wave functions, meaning that these states have strong-coupling cluster 
structures rather 
than weak-coupling ones.
In this strong-coupling cluster picture, the 
excitation from the $\Be(0^+_1)$ 
to the $\Be(0^+_2)$ is understood by the rotation of the 
deformed $\He$-cluster from the transverse configuration to the 
aligned configuration with some extent of spatial development of clustering.
Because of the difference in the $\He$-cluster orientation, i.e., the difference in two-neutron configurations, the ISM transition is suppressed for the transition 
between  $\Be(0^+_1)$ and $\Be(0^+_2)$ even though the $\Be(0^+_2)$
has the developed cluster structure.

The $\He+\alpha$ resonance states, $\Be(0^+_3)$ and $\Be(0^+_4)$, have
spatially developed cluster structures with significant cluster components at $D=6-7$ fm.
These states show intermediate features of 
strong-coupling and weak-coupling cluster structures. 
The $\Be(0^+_3)$ state has the dominant component of 
$\Phi^{(T)}_{\He+\alpha}(D)$ at $D=6-7$ fm
and is interpreted as a higher nodal state excited from the ground state in the model space 
of the transverse configuration (f). It has the 
significant $\He(0^+)+\alpha$ component and can be also 
regarded as a $\He(0^+)+\alpha$ resonance. 
The $\Be(0^+_4)$ has the 
relatively larger $\He(2^+)+\alpha$ component at $D=6-7$ fm and is regarded as 
a $\He(2^+)+\alpha$ resonance. 

Also in $\Li$, the bound states, 
$\Li(3/2^-_1)$ and  $\Li(3/2^-_2)$, can 
be clearly classified by the strong-coupling 
$\He+t$ wave functions.
The $\Li$ ground state has the dominant overlap with 
$\Phi^{(T)}_{\He+t}(D)$ at $D=2-3$ fm and almost no overlap with 
$\Phi^{(A)}_{\He+\alpha}(D)$, which is suppressed by the Pauli blocking
effect at the short distance. The $\Li(3/2^-_2)$ is dominated by $\Phi^{(I_z2)}_{\He+t}(D)$
at $D\sim 2$ fm and can be regarded as the $K=3/2$ state.
The resonance states of $\Li(3/2^-)$ have significant cluster components in 
the $D=5-7$ fm meaning the spatially developed $\He+t$ clustering. 
The $\Li(3/2^-_3)$ has the remarkable $\He(0^+)+t$ component and 
relatively small $\He(2^+)+t$ component, and therefore is 
regarded as a weak-coupling $\He(0^+)+t$ cluster resonance.
The $\Li(3/2^-_4)$ and $\Li(3/2^-_5)$ can be regarded as 
$\He(2^+)+t$  cluster resonances because they 
have significant $\He(2^+)+t$ components 
and relatively small $\He(0^+)+t$ components. 

The appearance of the weak-coupling $\He(0^+)+t$ cluster resonance in the $\Li(3/2^-_3)$
is a specific feature of the $\Li$ system different from the $\Be$ system, in which 
each $\He+\alpha$ cluster resonance is not a pure 
weak-coupling $\He(I^+)+\alpha$ state.
One of the keys for this difference between $\Li$ and $\Be$ 
is the presence or absence of the molecular $\sigma$-orbital structure
below the threshold energy.
In $\Be$, 
the molecular $\sigma$-orbital structure is favored and it appears
in the $\Be(0^+_2)$.  Note that 
the molecular $\sigma$-orbital structure corresponds to 
the $\He(A)+\alpha$ wave function in the present model.
It is important that the orthogonal condition of the $\Be(0^+_2)$ 
to the ground state is satisfied by the orthogonality of the 
orientation of the deformed $\He$ cluster. 
In higher states, the $\He+\alpha$ clustering develops keeping
the orthogonal condition to the lower states,  $\Be(0^+_1)$ and $\Be(0^+_2)$. 
The existence of the $\Be(0^+_2)$ having the aligned configuration $\He(A)+\alpha$ 
at the moderate distance ($D=4-5$ fm) somewhat 
suppresses the weak-coupling feature of $\He+\alpha$  cluster resonances
because the orthogonal condition to the $\Be(0^+_2)$ depends 
on the orientation of the deformed $\He$-cluster.
In contrast, the molecular $\sigma$-orbital structure is not favored in $\Li$
because of the asymmetry of the $\alpha+t$ core. 
In the developed cluster states, the rotational symmetry 
of the subsystem $\He$ is restored in the absence of the molecular $\sigma$-orbital structure,
and the weak-coupling 
$\He+t$ clustering is favored to form the $\He(0^+)+t$ cluster resonance in the $\Li(3/2^-_3)$.
 
\begin{figure}[th]
\begin{center}
\includegraphics[width=0.48\textwidth]{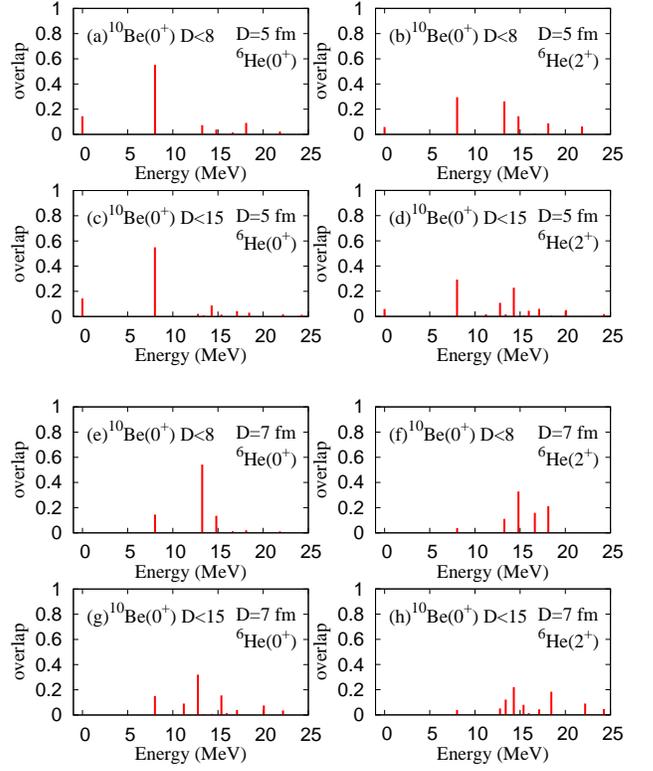} 	
\end{center}
\caption{Distributions of $\He(0^+)+\alpha$ and $\He(2^+)+\alpha$ components in 
$\Be(0^+)$. 
(a)(b) Squared overlaps of $\Be(0^+)$ obtained by the $D\le 8$ calculation
with $\Phi^{[0\otimes 0]_0}_{\He+\alpha}(D)$ and 
$\Phi^{[2\otimes 2]_0}_{\He+\alpha}(D)$ at $D=5$ fm, and 
(e)(f) those at $D=7$ fm. 
(c)(d)(g)(h) Same but for the $D\le 15$ calculation.
\label{fig:be10-norm}}
\end{figure}

The $\He(0^+)+\alpha$ and $\He(2^+)+\alpha$ components at $D=5$ fm and $D=7$ 
fm in $\Be$ are shown in Fig.~\ref{fig:be10-norm}.
The figure shows the squared overlaps of 
$\Phi^{[I\otimes J']_J}_{\He+\alpha}(D)$ with the $\Be(0^+)$
obtained by the $D\le 8$ and $D\le 15$ calculations.
At $D=5$ fm, the $\He(0^+)+\alpha$ component is concentrated at 8 MeV for 
the $\Be(0^+_2)$, which has also the significant $\He(2^+)+\alpha$ 
component showing the development of the strong-coupling cluster structure
(Fig.~\ref{fig:be10-norm}(a)-(d)).
At $D=7$ fm, the $\He(0^+)+\alpha$ component is concentrate on the 
$\Be(0^+_3)$ at 13 MeV, whereas the $\He(2^+)+\alpha$ component is significant 
in the $\Be(0^+_4)$ at 15 MeV in the $D\le 8$ calculation (Fig.~\ref{fig:be10-norm}(e) and (f)). Also in the $D\le 15$ calculations, the
concentration of the $\He(0^+)+\alpha$ component around 13 MeV and 
that of the $\He(2^+)+\alpha$ component around 15 MeV can be seen though 
the components are somewhat fragmented  (Fig.~\ref{fig:be10-norm}(g) and (h)).

\begin{figure}[th]
\begin{center}
\includegraphics[width=0.48\textwidth]{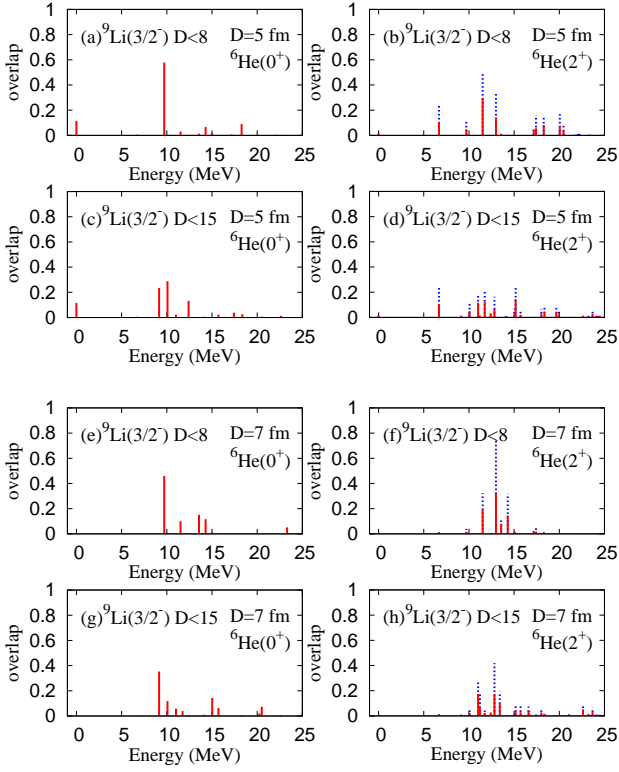} 	
\end{center}
\caption{Distributions of $\He(0^+)+t$ and $\He(2^+)+t$ components in 
$\Li(3/2^-)$. 
(a)(b) Squared overlaps of $\Li(3/2^-)$ obtained by the $D\le 8$ calculation
with $\Phi^{[0\otimes 3/2]_{3/2}}_{\He+t}(D)$ (solid lines), 
$\Phi^{[2\otimes 3/2]_{3/2}}_{\He+t}(D)$ (solid lines), and 
$\Phi^{[2\otimes 1/2]_{3/2}}_{\He+t}(D)$ (dashed lines) at $D=5$. 
(e)(f) Those at $D=7$ fm. 
(c)(d)(g)(h) Same but for the $D\le 15$ calculation.
}
\label{fig:li9-norm}
\end{figure}

Figure \ref{fig:li9-norm} shows 
$\He(0^+)+t$ and $\He(2^+)+t$ components  at $D=5$ fm and $D=7$ fm in $\Li$.
The distributions of the components are qualitatively similar at $D=5$ fm and 
$D=7$ fm.  The $\He(0^+)+t$ component is concentrated at 10 MeV for 
the $\Li(3/2^-_3)$  in the $D\le 8$ calculation (Fig.~\ref{fig:li9-norm}(a) and (e)).
Even in the $D\le 15$ calculation, the significant $\He(0^+)+t$ component  is found
in the corresponding energy region around 9 MeV (Fig.~\ref{fig:li9-norm}(c) and (g)).
The $\He(2^+)+t$ component is significantly contained in states in the 
$11-14$ MeV region.

It should be point out that 
the distributions of the $\He(0^+)+\alpha$ and $\He(0^+)+t$ components
shown in Figs.~\ref{fig:be10-norm} and  \ref{fig:li9-norm} are not necessarily 
consistent with the ISM strength distributions shown in 
Figs.~\ref{fig:be10-is0} and \ref{fig:li9-is0}. 
In particular, in spite of the remarkable $\He(0^+)+t$ component 
in the $\Li(3/2^-)$ state around 10 MeV, there is no significant 
ISM strength in the corresponding energy region.  Also for  $\Be(0^+)$, 
even though the $\Be(0^+_2)$ has the remarkable  $\He(0^+)+\alpha$ 
component, the ISM transition to this state is 
relatively suppressed compared with other states.
As discussed later, there is no one to one correspondence between 
the ISM excitation and the $\He(0^+)+\alpha$ component, but 
the ISM operator more directly excites the specific type of strong-coupling 
cluster structures embedded in the ground state. 

\subsection{Isoscalar monopole excitations in cluster structures}

Enhancement of ISM transitions to cluster states 
has been discussed for stable nuclei and also unstable nuclei 
\cite{Ito:2011zza,Suzuki:1989zza,Kawabata:2005ta,Yamada:2011ri}. 
As discussed by Yamada {\it et al.}, ISM strengths are enhanced 
for transitions from the ground state to cluster excited states because 
the ISM operator ${\cal M}(IS0)$ excites the inter-cluster motion 
through the $r^2$ term in ${\cal M}(IS0)$ \cite{Yamada:2011ri}.
As shown in Ref.~\cite{Yamada:2011ri}, ${\cal M}(IS0)$ can be rewritten as 
\begin{equation}
{\cal M}(IS0)=\frac{A_1A_2}{A}r^2+\sum_{i \in C}\left(\bvec{r}_i-\bvec{R}_1\right)^2+
\sum_{i \in C'}\left(\bvec{r}_i-\bvec{R}_2\right)^2,
\end{equation}
where $\bvec{R}_1$ and $\bvec{R}_2$ is the center of mass coordinates of the first
($C$) and second ($C'$) clusters. 
If the antisymmetrization is ignored, the first term excites the inter-cluster motion
whereas the second and third terms cause the ISM excitations 
of $C$ and $C'$ clusters. 

In the ISM excitations, the ${\cal M}(IS0)$ operated state, 
${\cal M}(IS0)|\textrm{g.s.}\rangle$ is regarded as the door-way state
that is initially produced by the ISM excitation. 
Provided that excitations of the inter-cluster motion
(cluster mode) are decoupled well from internal excitations of the clusters, 
the ISM strengths to cluster states are nothing but
distributions of the door-way state ${\cal M}(IS0)|\textrm{g.s.}\rangle$
projected onto the cluster model space, 
$\hat P(C+C'){\cal M}(IS0)|\textrm{g.s.}\rangle$, which 
approximately corresponds to the door-way state excited from the ground state by 
the $r^2$ operator. Here $\hat P(C+C')$ is the projection operator onto the 
$C+C'$ cluster model space. 
In the present case, ISM
strengths indicate 
the distributions of the projected 
door-way state $\hat P(\He+\alpha(t)){\cal M}(IS0)|\textrm{g.s.}\rangle$
in the $\He+\alpha(t)$ cluster model space.

\begin{figure}[th]
\begin{center}
\includegraphics[width=7cm]{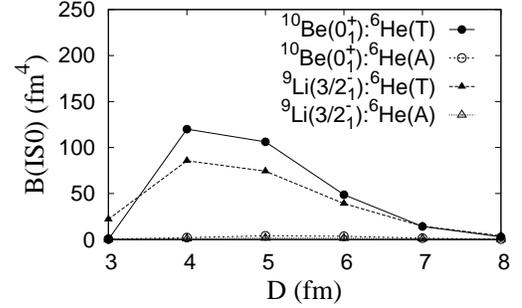} 	
\end{center}
\caption{ISM strengths 
$B(IS0;\textrm{g.s.}\to \Phi^{(T,A)}_{\He+\alpha(t)}(D))$
for the transitions from the $\Be$ and $\Li$ 
ground states obtained by the $D\le 8$ calculation to
$\Phi^{(T,A)}_{\He+\alpha(t)}(D)$.
$B(IS0;\textrm{g.s.}\to \Phi^{(T,A)}_{\He+\alpha(t)}(D))$
 is defined in \eqref{eq:bis0-base}.
}
\label{fig:bis0-base}
\end{figure}

As discussed previously, the low-lying states of $\Be$ and $\Li$ are 
understood by strong-coupling cluster structures. In particular, the
ground state is dominated by the transverse configuration
$\Phi^{(T)}_{\He+\alpha(t)}(D)$ 
with the distance $D=2-3$ fm. Therefore, the projected 
door-way state $P(\He+\alpha(t)){\cal M}(IS0)|\textrm{g.s.}\rangle$
is approximately included by the subspace with the specific configuration
$\Phi^{(T)}_{\He+\alpha(t)}(D)$ because the ISM operator excites 
the inter-cluster motion through the 
$r^2$ term but does not change the orientation of the deformed $\He$ cluster.
Figure \ref{fig:bis0-base} shows the ISM strengths for transition from the 
$\Be$ and $\Li$ ground state to  
specific configurations. One is the transverse configuration  $\Phi^{(T)}_{\He+\alpha(t)}(D)$ and the other is the aligned configuration
$\Phi^{(A)}_{\He+\alpha(t)}(D)$. The strengths of the ISM transitions from the 
ground state is calculated as
\begin{eqnarray}
&&B(IS0;\textrm{g.s.}\to \Phi^{(T,A)}_{\He+\alpha(t)}(D))\nonumber \\
&&=\frac{1}{2J+1}
\left|\left\langle\textrm{g.s.}  ||{\cal M}(IS0)||\Lambda_\textrm{g.s.}\Phi^{(T,A)}_{\He+\alpha(t)}(D)\right\rangle \right|^2, \label{eq:bis0-base}\\
&&\Lambda_\textrm{g.s.}\equiv 1-|\textrm{g.s.}\rangle \langle \textrm{g.s.}|,
\end{eqnarray}
where the normalizations of the initial and final states are chosen to be one, 
and the orthogonal condition of the final state to the initial state is satisfied by the 
projection operator $\Lambda_\textrm{g.s.}$.
As seen in the figure, the calculated ISM strengths show 
remarkable transitions to $\Phi^{(T)}_{\He+\alpha(t)}(D)$ at $D=4-5$ fm
but almost no transition to $\Phi^{(A)}_{\He+\alpha(t)}(D)$
as expected from the dominant $\Phi^{(T)}_{\He+\alpha(t)}(D)$ component 
in the initial state $|\textrm{g.s.}\rangle$. 
It means that the door-way state excited from the ground state 
by the ISM operator dominantly contains 
the transverse configuration of the $\He$ cluster. 

\begin{figure}[th]
\begin{center}
\includegraphics[width=0.48\textwidth]{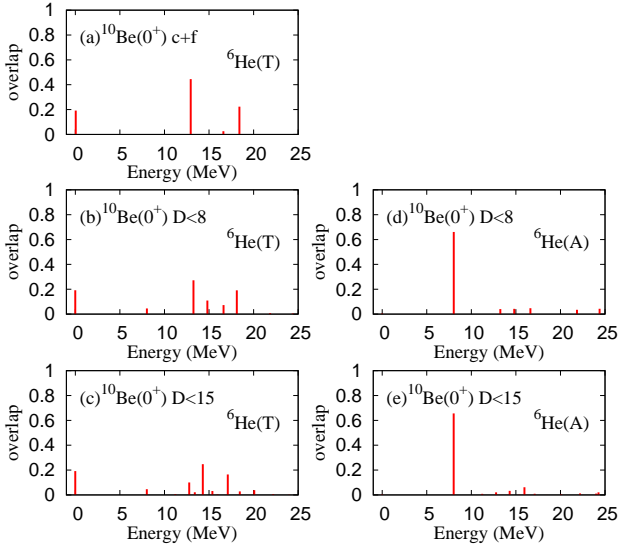} 	
\end{center}
\caption{Distributions of squared overlaps of  
$\Phi^{(T,A)}_{\He+\alpha}(D)$ at $D=5$ fm
with the $\Be(0^+)$ obtained by the (c+f), $D\le 8$, and $D\le 15$ calculations.
}
\label{fig:be10-norm-1b}
\end{figure}

\begin{figure}[th]
\begin{center}
\includegraphics[width=0.48\textwidth]{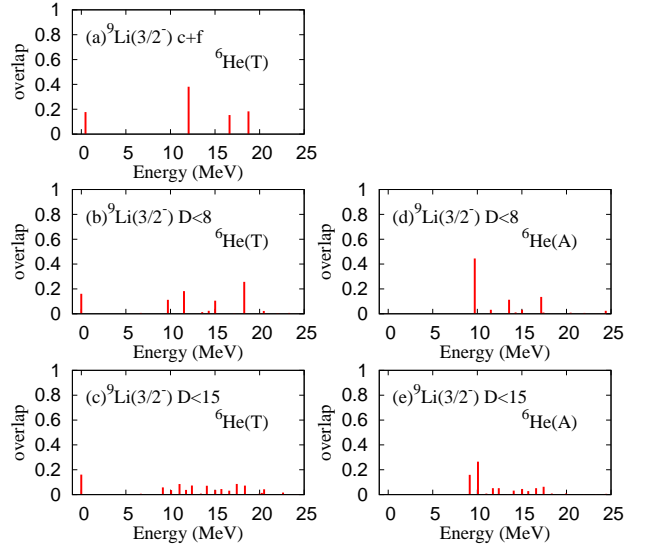} 	
\end{center}
\caption{Distributions of squared overlaps of  
$\Phi^{(T,A)}_{\He+t}(D)$ at $D=5$ fm
with the $\Li(3/2^-$) obtained by the (c+f), $D\le 8$, and $D\le 15$ calculations.
}
\label{fig:li9-norm-1b}
\end{figure}

Let us discuss again the IMS strengths obtained by three 
calculations, (c+f), $D\le 8$, and $D\le 15$.
As already shown in Table \ref{tab:is0}, 
the EWS obtained by the (c+f) calculation 
is almost consistent with that obtained by the 
$D\le 8$ calculation with full $\He$ configurations.
It meant that the door-way state directly 
produced from the ground state with the ISM operator are mostly
contained in the truncated model space of the configurations (c) and (f).
As shown in Figs.~\ref{fig:be10-is0}(a) and \ref{fig:li9-is0}(a), 
the ISM strengths are concentrated in a few low-lying states in the 
the (c+f) calculation meaning that the door-way state is 
distributed in these few states of the truncated model space. 
However, in the $D\le 8$ calculation with full configurations of 
$\He+\alpha(t)$, the ISM strengths are somewhat fragmented 
through coupling of the configuration $\Phi^{(T)}_{\He+\alpha(t)}(D)$
with other configurations. The ISM strengths 
should reflect the distributions of the $\Phi^{(T)}_{\He+\alpha}(D)$ 
and $\Phi^{(T)}_{\He+t}(D)$ 
components in the obtained $\Be(0^+)$ and $\Li(3/2^-)$.
As already shown in Fig.~\ref{fig:bis0-base}, 
the $\Phi^{(T)}_{\He+\alpha(t)}(D)$ configuration has the strong ISM transition 
in the $D=4-5$ fm region meaning that the door-way state
excited from the ground state by the ISM operator has large overlap with this state.
In Figs.~\ref{fig:be10-norm-1b} and \ref{fig:li9-norm-1b}, I show the squared overlaps of $\Be(0^+)$ and $\Li(3/2^-)$ obtained by three calculations with 
$\Phi^{(T)}_{\He+\alpha}(D)$ and $\Phi^{(T)}_{\He+t}(D)$ at $D=5$ fm.
For comparison, I also show the components of the aligned configuration $\Phi^{(A)}_{\He+\alpha(t)}(D)$ at $D=5$ fm.
Comparing the results of Figs.~\ref{fig:be10-is0} and \ref{fig:li9-is0} and 
those of Figs.~\ref{fig:be10-norm-1b} and \ref{fig:li9-norm-1b}, 
the ISM strength distributions can be qualitatively described by the  
distributions of the $\Phi^{(T)}_{\He+\alpha(t)}(D)$ component. 
In the $D\le 15$ calculation of $\Li$,  
the fragmentation of the ISM strengths for the $\He+t$ cluster 
resonances around 10 MeV is understood by the strong fragmentation 
of the $\Phi^{(T)}_{\He+\alpha(t)}(D)$ component because of the 
coupling with other configurations.

\begin{figure}[th]
\begin{center}
\includegraphics[width=7cm]{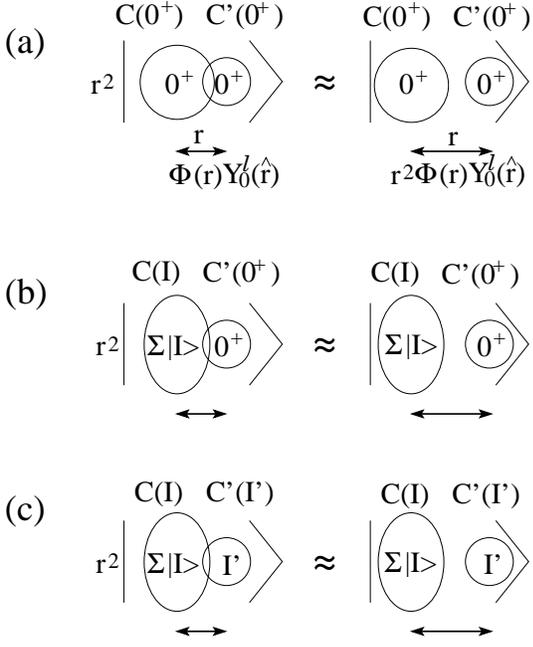} 	
\end{center}
\caption{Schematic figures for ISM excitations to cluster states.
(a) A system of two clusters that are spin-less $ls$-closed shell nuclei. 
(b) A system of a deformed cluster and a spin-less $ls$-closed  cluster.
(b) A system of a deformed cluster and a finite-spin cluster.
}
\label{fig:c1-c2}
\end{figure}

Let us give more general discussions of the door-way state excited from the ground state
to cluster states by the ISM operator. As mentioned previously, the projected door-way state approximately corresponds to the $r^2$ operated state of the 
ground state.  In case that the ground state contains a deformed cluster with a specific 
orientation, the inter-cluster motion is excited from the ground state to 
the door-way state
keeping the orientation of the cluster as it is in the ground state
because the $r^2$ operator does not bring rotation of clusters.
Schematic figures are shown in Fig.~\ref{fig:c1-c2}.
When the system consists of two spin-less clusters such as 
$^{16}{\rm O}+\alpha$, in which both clusters are $ls$-closed shell nuclei,  
the projected door-way state can be concentrated 
on a $^{16}{\rm O}+\alpha$ cluster state, and therefore,  
the ISM transition should be strong for the
$^{16}{\rm O}+\alpha$ cluster state (see Fig.~\ref{fig:c1-c2}(a)). 
However, in the cases of $\He+\alpha$ and $\He+t$ cluster states in 
$\Be$ and $\Li$, the $\He$ cluster is not a $ls$-closed shell nucleus but
is deformed in the $\Be$ and $\Li$ ground states 
because of the Pauli blocking effect between clusters  
(see Fig.~\ref{fig:c1-c2}(b) and (c)).
The deformed $\He$ cluster is the mixed state of different 
spin states, $\He(I)$. In the asymptotic region of a large inter-cluster distance, 
the deformed cluster is not favored but the angular-momentum eigen states
$\He(I)$ are favored because of the restoration 
of the rotational symmetry of the subsystem.
Moreover, in the $\He+t$ system,
the second cluster ($t$) has the finite intrinsic spin $I'=1/2$. 
Therefore, the projected door-way state is fragmented in
$\He+t$ cluster states through the coupling 
of the angular momenta $I$ and $I'$ of the clusters 
and the orbital angular momentum of the inter-cluster motion. 
As a result, the ISM strengths can be strongly fragmented.
In the $\He+\alpha$ system, since 
the angular momentum coupling is not so strong for the spin-less $\alpha$ cluster, the fragmentation of the door-way state can be weaker than the $\He+t$ system. Therefore, the fragmentation of the 
ISM strengths to $\He+\alpha$ cluster states is 
not so strong as that to $\He+t$ cluster states.

\section{Summary}\label{sec:summary}

The ISM transitions from the ground states 
to cluster states in $\Be$ and $\Li$ 
were investigated with the $\He+\alpha$ and $\He+t$ cluster models,
respectively.
In the calculation, the resonance states were obtained in a bound state approximation. 
The $\He+\alpha$ and $\He+t$ widths of the resonances were estimated by the method
using the reduced width amplitudes obtained by the 
calculation with $D\le 8$ fm. 
The coupling with continuum states evaluated  
by changing the boundary size from $D\le 8$ fm to  $D\le 15$ fm
shows consistent results with the decay widths estimated by the bound 
state approximation with $D\le 8$ fm.

In $\Be$, the $\He+\alpha$ cluster resonances 
were obtained as the $\Be(0^+_{3})$ and $\Be(0^+_{4})$ 
above the $\He+\alpha$  threshold energy. 
The significant ISM strengths were obtained for the transitions to 
these resonances.
In $\Li$, the $\He+t$ cluster resonances
were obtained  as the $\Li(3/2^-_{3,4,5}) $ above the $\He+t$ threshold energy. 
The ISM strengths are strongly fragmented and 
show no enhancement for these $\He+t$ cluster resonances
differently from the $\He+\alpha$ cluster resonances in $\Be$.

By analyzing the cluster 
components of excited states, 
the relation of the ISM excitations with the cluster components 
were discussed.
It was found that 
the ISM strength distributions do not directly 
correspond to the distributions of the
$\He(0^+)+\alpha$ and $\He(0+)+t$ components but 
they are distributed via components of the deformed $\He$
cluster configuration with a specific orientation. 
The ISM strengths to cluster states are the
distributions of the door-way state, which is excited from the ground state by 
the $r^2$ operator, in the cluster model space. 
Since the $\Be$ and $\Li$ ground states are dominated by the 
the transverse configuration of the deformed $\He$ cluster, 
the door-way state also dominantly contains the transverse configuration
because the ISM operator excites 
the inter-cluster motion through the 
$r^2$ term but does not change the orientation of the deformed $\He$ cluster.
The door-way state distributions, which is originally concentrated 
on the specific configuration, are fragmented in final states in the full model space 
because of mixing with other configurations 
as well as the angular momentum coupling. 
This is an interpretation of the fragmentation of the ISM strengths in $\Be$ and $\Li$.
It should be stressed that the ISM excitations more directly reflect the 
strong-coupling cluster features which is originally embedded in the ground state
rather than the weak-coupling cluster features. 

\section*{Acknowledgments}
The author would like to thank Dr.~Kimura
and Dr. Suhara for fruitful discussions.
The computational calculations of this work were performed by using the
supercomputer in the Yukawa Institute for theoretical physics, Kyoto University. This work was supported by 
JSPS KAKENHI Grant Number 26400270.

\appendix
\section{Calculation of partial decay widths in the bound state approximation}\label{app:width}
In the default $D\le 8$ calculation, the resonance states are obtained 
as bound state solutions 
in the model space of $D \le 8$ fm.
In a bound state approximation, 
the partial decay width $\Gamma_{I\otimes J'}$
of a resonance state for $\He(I^+)+\alpha(t)$ channels 
with the angular momentum coupling $[I\otimes J']_{J}$ can be estimated
from the reduced width amplitude
$y(a)$ of the corresponding channel at a channel radius $a$, 
\begin{eqnarray}
\Gamma_{I\otimes J'}&=&\frac{2ka}{F^2_l(ka)+G^2_l(ka)}\gamma^2(a),\\
\gamma^2(a)&=&\frac{\hbar^2}{2\mu a}\left[ ay(a) \right]^2,
\end{eqnarray}
where $F_l$ and $G_l$ are the regular and irregular Coulomb functions, respectively, 
$k$ is the momentum of inter-cluster motion in the asymptotic region, and 
$\mu$ is the reduced mass. $\gamma^2(a)$ is the so-called reduced width. 
$l$ is the orbital angular momentum of the relative motion. 
In the present work, $y(a)$ are 
approximately calculated by using the overlap with the $\He(I^+)+\alpha(t)$
cluster wave function $\Phi^{[I\otimes J']_J}_{\He+\alpha(t)}(D=a)$ by means of the method proposed in Ref.~\cite{Kanada-En'yo:2014nla} as,
\begin{eqnarray}\label{eq:y-app}
ay(a) \approx
\frac{1}{\sqrt{2}}\left(\frac{2\gamma}{\pi}\right)^{1/4}
\left\langle \Psi^{J^\pi_k}| \Phi^{[I\otimes J']_J}_{\He+\alpha(t)}(D=a)\right\rangle.
\end{eqnarray}

\end{document}